\documentclass[12pt]{article}

\usepackage{spie_jnp}

\begin{document}

\mytitle{Propagation of localized surface plasmons in sets of
metallic nanocylinders at the exit of subwavelength slits}

\myauthor{F.J. Valdivia-Valero$^{1}$ and M. Nieto-Vesperinas$^{1}$}

\myaddress{$^{1}$Instituto de Ciencia de Materiales de Madrid,
C.S.I.C., Campus de Cantoblanco \\ 28049 Madrid, Spain}

\myemail{fvaldivia@icmm.csic.es; mnieto@icmm.csic.es}

We analyze, by means of numerical simulations, transmission
enhancements through subwavelength slits due to the presence of sets
of plasmonic nanocylinders, placed near the exit of these apertures.
Further, we extend this study to  photonic crystals of dipolar
plasmonic particles in front of an array of extraordinarily
transmitting slits practiced in a metallic slab.

\keywordline (050.1940) Diffraction; (050.1220) Apertures;
(050.6624) Subwavelength structures; (160.4236)  Nanomaterials;
(230.5750) Resonators; (230.7370) Waveguides; (240.6680) Surface
plasmons; (250.6715) Switching.

\section{Introduction}
\label{}

\emph{Localized surface plasmons} (LSP) of metal nanoparticles \cite{Zayats1},
\cite{Sburlan06}, \cite{Zayats2}, \cite{Bryant08} enjoy a potential as elements of
nanooptical networks. These resonant modes arise from the excitation
of coherent oscillations of conduction-band electrons, localized on
the surface of these particles. The strong light coupling leads to
its absorption and spatial confinement to a nanometric scale, which
results in large local enhancements of electromagnetic field
intensities. Ensembles of particle chains have been extensively
studied \cite{Maier}. That provides nanoscale control of the
transmission, manipulation and switching of optical signals
\cite{Maier2}. In addition, nanoparticles either dielectric
\cite{Astratov} or metallic \cite{Tesis} have been studied in
microdisks in connection with the mutual perturbation of their
resonances, such as counterpropagating \emph{whispering gallery
mode} (WGM) splitting, modification and their monitoring in the
cavity, or control of the radiating properties of the nanoparticle
by the WGMs of the cavity \cite{Mazzei}.

A somewhat related phenomenon known as enhanced optical transmission
through subwavelength apertures, either alone or forming a grating,
\cite{Ebbesen98}, \cite{Garc�a-Vidal02}, \cite{Porto99},
\cite{Lezec04}, \cite{Garc�a07}, has received much attention in
connection with its potential application for light concentration,
detection and wavefront steering.

These morphology dependent  resonances (MDR) also play an important
role in  \emph{photonic crystals} (PC). For example, their influence
on the crystal bandgap size and position has been studied
\cite{Soukoulis98}, \cite{Moroz99}. In this way, the Mie resonances
of the particles forming the  PC  constitute the light propagation
and enhancement vehicle in the upper bands of those so called {\em
molecular photonic crystals} \cite{Vigneron}. On the other hand, as
regards metallic PCs, there are certain advantages in fabricating
them, like reduced size and weight, easier production methods and
lower costs, as well as the fact that low loss metal PCs have been
studied \cite{Soukoulis00}. Furthermore, many applications are being
developed from these structures, like waveguiding \cite{He00},
\cite{Zhao07}, \cite{Wen03}, \cite{Al_Naboulsi04}, waveguide
mode-plasmon coupling \cite{Giessen03}, light transmission control
\cite{Fan05}, thermovoltaics and blackbody emission \cite{Ho02},
\cite{Gu10}, and lensing \cite{Zavada07}.

In this paper we wish to answer two questions: first, what is the
effect of the presence of plasmonic particles in the transmission
zone of a subwavelength aperture which is supertransmitting? Do
these particles enhance or inhibit this supertransmission?. Second,
how does the energy flow of transmitted light by this slit, (or
slits), propagate through these sets of particles?. We present a
study by means of numerical simulations that show new effects in
configurations of plasmonic nanoparticles near the exit of
supertransmitting subwavelength slits practiced in a thick slab. Our
study is carried out in 2D but the essential features observed as
regards enhanced transmission and coupling of resonances and light
transport are likewise obtained in 3D \cite{Taflove1}
\cite{VandeHulst}. Also, this 2D geometry constitutes a good model
with equivalent effective constitutive parameters for microdisks
\cite{Chin}, \cite{Boriskina06_2}. Furthermore, such a geometry is
adequate to deal with structures of long parallel nanocylinders in
2D PCs or metamaterials \cite{Vynk}. We then address localized
surface plasmons of metallic nanocylinders \cite{Arias-G}. We see
the behavior of light concentration and transmitted field
enhancement efficiency in this kind of particle sets (which appears
maximized when the configuration of the plasmonic set has properties
similar to those of a nanoantenna).

Further calculations deal with configurations  like  linear and
bifurcated nanocylinder chains, also addressing the natural step
passing from particle chains, placed one close to another, towards a
\emph{photonic crystal (PC)} geometry situated in front of a
metallic array of slits. In this way, we design a method of
collimation and coupling of light from free space into the particles
by placing them close to the slits. These numerical simulations
allow to study the effects that arise in the near field as regards
extraordinary transmission in the slits and the excitation of LSPs
of these metallic particles; and since the calculations are exact,
they constitute a reliable design of future experiments that can be
performed in either 2D or 3D. In this respect, our calculations
indicate that whereas the nanoaperture behaves as a collimator that
transmits light into the nanoparticles, the fundamental role as
regards enhancement and field concentration, corresponds to the
excitation of the particle LSP which couple with the MDRs of the
slits. This is thoroughly studied by first displaying the simple
configuration of a single plasmonic nanoparticle in front of a
nanoaperture, and afterwards extending these observations to more
complex sets of these particles, including chains and PCs. We take
advantage, both in chains and in PC geometries, of the excitation of
dipolar LSPs.

\section{Transmission into plasmonic nanoparticles through a nanoslit}

\subsection{Numerical procedures.}

From now on, all refractive indexes under the different wavelengths
on use are taken from \cite{Johnson} and \cite{Palik}. All particles
in this study are considered of Silver ($Ag$), (refractive index $n=
0.188+i1.610$ at $\lambda= 366nm$ and $n= 0.233+i1.27$ at $\lambda=
346nm$) because of their rich LSP spectrum in the near ultraviolet.
However, it should be stressed here that this is done for the sake
of illustrating the effects, and that other noble metal material can
be chosen. Since the slab is thicker than usual in supertransmission
experiments, in order that the slit walls present high reflection
and as small as possible skin depth and losses, the metal of the
slab is assumed to be Tungsten ($W$) (refractive index $n=
3.40+i2.65$ at $\lambda= 366nm$ and $n= 3.15+i2.68$ at $\lambda=
346nm$). It should be remarked in this connection that, ideally, a
quasi-perfect conducting slab would exhibit the most pronounced
effects under study, but if experiments are done with thinner slabs
these other materials like $Al$, $Au$ or $Ag$ may be employed. The
conservation of light polarization is ensured by launching on the
aperture linearly polarized light beams of rectangular profile,
their widths being that of the simulation window (this one always
coincides to the slab width). The direction of propagation of such
beams is normally incident to the axis of the infinite cylinders,
and their sense is, in all cases, that from the bottom to the top of
the window. The wavelengths utilized to visually show the slits as
well as the particle - slit configurations have been chosen so as to
maximize their light transmission. The distances between all of such
particle structures and the plane exit of the slit, optimizing their
responses, have been expressed in semi-multiples of the
corresponding LSP resonance - to - supertransmission matched
wavelength (with the wavelength resolution limited by \cite{Palik}).
Maxwell equations are solved by using a finite element method (FE)
(FEMLAB of COMSOL, \mbox{http://www.comsol.com}) \cite{Caloz},
\cite{Garcia-Pomar04}. The solution domain is meshed with element
growth rate: 1.55, meshing curvature factor: 0.65, approximately;
the geometrical resolution parameters consist of 25 points per
boundary segment to take into account curved geometries so as to
adapt the finite elements to the geometry and optimize the
convergence of the solution. The final mesh contains about $10^{4}$
elements. To solve Helmholtz equation, the UMFPACK direct is
employed, the results being displayed in stationary regime of
propagation. The boundary conditions of the simulation space are
properly set both to keep the calculations from undesired window
reflections and to avoid possible geometrical discontinuities. We
hence ensure that neither inconsistencies due to properties
discontinuities of the objets under study nor possible systematic
errors because of the simulation window interfere with the field
calculation. {\it We always select p - polarized incident waves to
seek extraordinary transmission in this 2D slit} (s - polarized
waves do not produce such a phenomenon in 2D subwavelength slits)
\cite{Porto99}, \cite{Garc�a07}, \cite{Garc�a06}. The results
are thus expressed in terms of either the magnetic vector ${\bf
H}({\bf r}) [A/m]$ which is along the cylinder OZ - axis, the
electric vector ${\bf E}({\bf r}) [V/m]$ and the time average energy
flow $<{\bf S}({\bf r})> [J/(m^{2}\cdot s)]$, these last two being
both transversal, namely, in the plane of the images to show next.
The incident wave is a tapered rectangular plain wave of unit
amplitude in $|{\bf H}|= 1A/m$(SI), this value corresponding to
$<{\bf S}> \approx 190W/(m^{2})$. Finally, the nomenclature
followed, further on, to classify the LSP resonances of the
cylinders will use the subscripts \emph{(i, j)}, \emph{i} and
\emph{j} standing for their angular \emph{i - th} and radial \emph{j
- th} orders, respectively.

\subsection{One metallic particle in front of a nanoslit}

\begin{figure}[htbp]
\centering
\includegraphics[width=16cm]{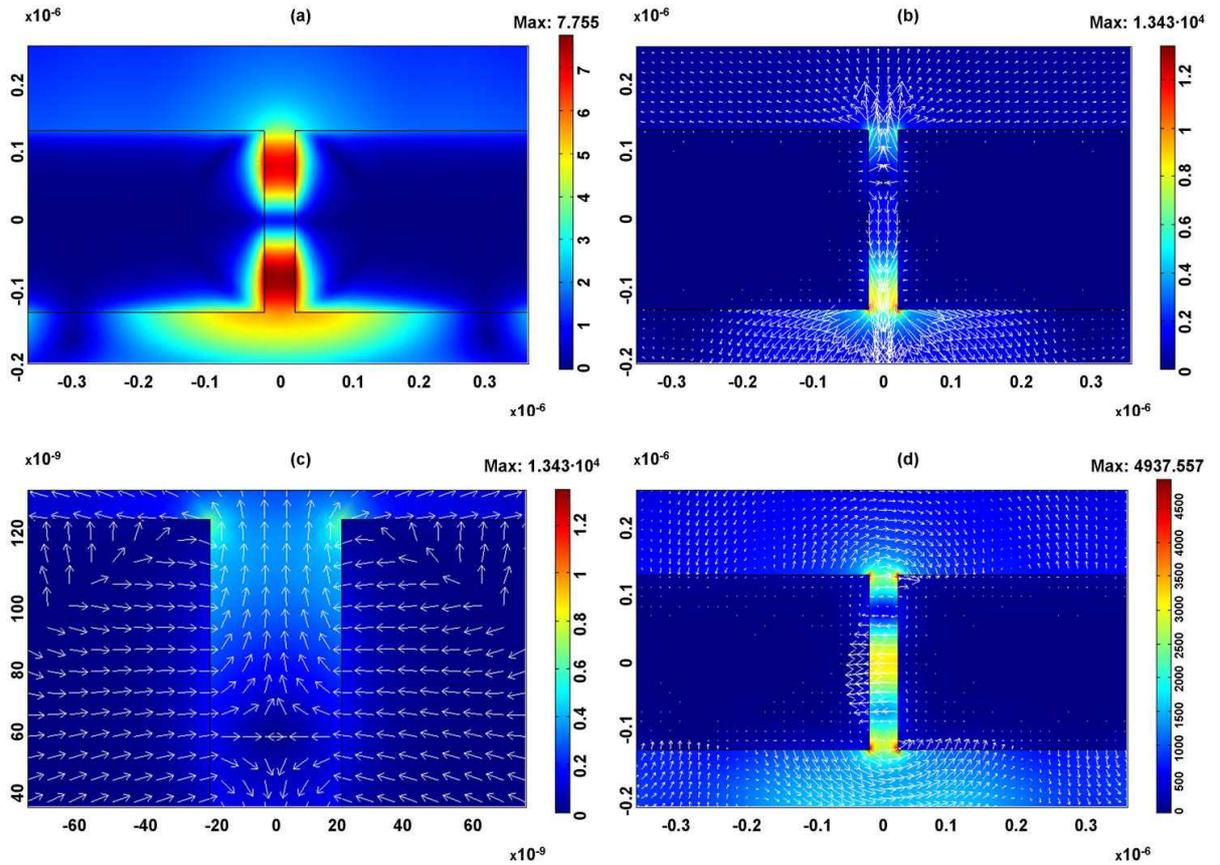}
\caption{(a) Magnetic field modulus $|{\bf H}({\bf r})|$ (in $A/m$
units), in a W slab aperture (refractive index $n= 3.39+i2.66$, slab
width $D= 2850nm$, slab thickness $h= 259.40nm$, slit width $d=
43.23nm$). The magnetic vector of the incident radiation ($\lambda=
364.7nm$) is p - polarized, of unit amplitude and impinges on the
slab from below; (b) Time average energy flow $<{\bf S}({\bf r})>$
(in $J/(m^{2}\cdot s$ units), maximum arrow length $\approx 83.82
KeV/(nm^{2}\cdot s)$, minimum arrow length $\approx 0eV/(nm^{2}\cdot
s)$), both norm (colors) and directions (arrows) are shown under the
same conditions as in Fig. 1(a); (c) Detail of the saddle point
evidenced by $<{\bf S}({\bf r})>$ inside the slit (the arrows here
appear normalized to their magnitude which is shown according to the
color bar); (d) Detail of the electric field {\bf E}({\bf r}) (in
{\it V/m} units), both its norm (colors) and directions (arrows) are
shown under the same conditions as in Fig. 1(a).}
\end{figure}

The phenomenon of field transmission enhancement through a
subwavelength aperture by excitation of LSP resonances in nearby
plasmonic particles, is studied by comparing it with the
transmission through a slit alone,  practiced in a metallic slab. We
chose $\lambda= 364.7nm$ for the incident wave, where the aperture
is supertransmitting. Figure 1(a) as well as Fig. 1(b) show the
magnetic field norm $|{\bf H}({\bf r})|$ and the time averaged
energy flow $<{\bf S}({\bf r})>$ distributions in such a slit,
respectively (in all figures, from now on, the p - polarized wave
illumination incides upwards). These distributions, which correspond
to an incident wavelength at which the slit produces extraordinary
transmission, show two interesting features: first, the waveguide p
- eigenmode inside the aperture and, second, the change of direction
of the energy flow, from the lower region of the aperture at which
the energy is partly reflected, to the upper region of the slit
where the energy manifests transmission upwards. This sign change is
evidenced in Fig. 1(c) by the energy flow as a potential saddle
point due to a change of sign of the magnetic field ${\bf H}({\bf
r})$. On the other hand, Fig. 1(d) exhibits the electric field ${\bf
E}({\bf r})$ which shows a strong charge concentration, and a
resulting a dipolar pattern configuration at the corners of both the
entrance and the exit of the slit (see the two upper and lower
vertices of the aperture in this figure). Incidentally, we believe
that this high intensity concentration at the edges of the aperture
exit is responsible for the existence of gradient optical forces on
dielectric nanoparticles placed in its proximity, and the
corresponding creation of two potential wells, each of them in front
of each  corner of the aperture exit [cf. Fig. 1(d)], as observed in
\cite{Quidant}. Notice in those figures that the supertransmitted
intensity by the slit is already seen in the spatial distribution of
$|{\bf H(r)}|$, ${\bf E(r)}$ and $|<{\bf S(r)}>|$ in the
neighborhood of the exit of the slit, (we recall that incident
$|{\bf H}|= 1A/m$, $|{\bf E}|= 380V/m$ and $|<{\bf S}>|=
190W/(m^{2}$).

\begin{figure}[htbp]
\centering
\includegraphics[width=16cm]{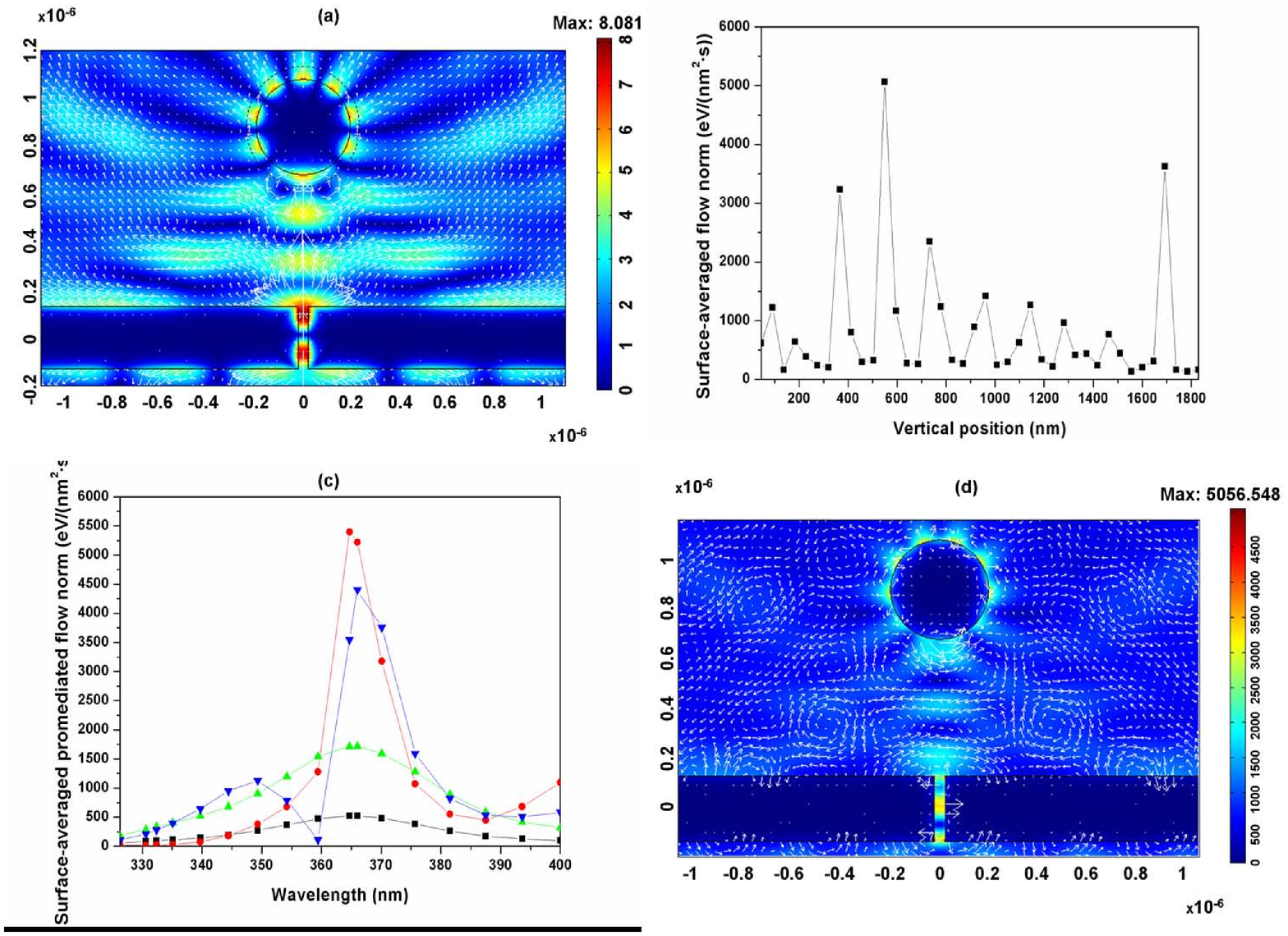}
\caption{(a) Magnetic field modulus $|{\bf H(r)}|$ (colors in $A/m$
units) and time averaged energy flow $<{\bf S(r)}>$, (arrows in
$J/(m^{2}\cdot s$ units), maximum arrow length $\approx 80.02
KeV/(nm^{2}\cdot s)$, minimum arrow length $\approx 0eV/(nm^{2}\cdot
s)$), localized on the surface of an $Ag$ cylinder (radius $R=
200nm$, refractive index $n= 0.186+i1.61$) in front of a slit in a
$W$ slab at the same illumination as in Fig. 1(a). The distance
between the cylinder surface and the exit plane of the slit is
$d_{lc}= 3\lambda/2= 549nm$ ($\lambda= 366nm$); (b) Variation of the
concentration of $|<{\bf S(r)}>|$, (in $eV/(nm^{2}\cdot s$ units),
on the cylinder surface vs its vertical separation (in {\it nm})
from the slit exit, illuminated at $\lambda= 366nm$ (near the
plasmon $LSP_{51}$ resonance); (c) The same quantity vs wavelength
at the distance seen in Fig. 2(a), $d_{lc}= 549nm$. The black and
green curves stand for the response of the slab alone and the red
and blue ones for the slab with cylinder, respectively. (d) Electric
field {\bf E}({\bf r}) (in {\it V/m} units); both its norm (colors)
and directions (arrows), under the same conditions as in Fig. 2(a).
Black and red curves in Figs. 2(b) and 2(c) have been calculated by
averaging the quantity in an annulus of area $A= \pi
(((9/8)r)^{2}-r^{2})$ (see the two concentric circles in (a)) whose
internal circle coincides with either the cylinder section, (case of
the slit with the particle in Figs. 2(b) and 2(c)) or an imaginary
circle coincident with that cylinder section, (case of the slit
alone in Fig. 2(c)). Green and blue curves in Fig. 2(c) have been
calculated by averaging the quantity in a rectangular monitor of
area $S= 76nm x 58.46nm= 4442.96nm^{2}$ at the slit exit, (see the
rectangle drawn in Fig. 2(a)).}
\end{figure}

As a first example of the effect placing a plasmonic particle near
the aperture exit, Fig. 2(a) deals with an $Ag$ nanocylinder placed
in front of the exit of a subwavelength slit in a W slab, which
presents a large transmission at the chosen wavelength. As shown, at
$\lambda= 366nm$ and at a distance between the cylinder and the slab
$d= 549nm$, a strong intensity enhancement of $|{\bf H}({\bf r})|$
appears on the cylinder surface, manifested by the LSP stationary
interference pattern surrounding it, and a standing wave pattern due
to reflections between the slab/aperture and the particle is
observed. It should be noticed that this cylinder alone has this
$LSP_{51}$ mode at $\lambda= 359.7nm$ which is red - shifted as seen
in Figs. 2(a), 2(c) and 2(d) in presence of the slab and aperture. A
well - known phenomenon explained on the basis of the driven
oscillator model \cite{Sburlan06}. Also, both vortices and a saddle
point are exhibited by $<{\bf S}({\bf r})>$ in the standing wave
pattern between the nanocylinder and the slab. The vertical
separation between the cylinder and the slit exit has been chosen in
order to optimize this field enhancement. Figure 2(b) shows this
field concentration in terms of the energy flow magnitude $|<{\bf
S}>|$ as the nanocylinder is gradually moved away from the slit at
$\lambda= 366nm$. In order to quantify this, we have averaged this
quantity in an annulus surrounding the cylinder, whose internal
circle coincides with the cylinder section. The image of Fig. 2(a)
corresponds to the third peak of Fig. 2(b) from the left. The
spectrum of this enhancement vs the wavelength with the cylinder at
$d= 549nm$, can be seen in Fig. 2(c), which shows a comparison
between this enhancement with the cylinder at the aforementioned
distance $d= 549nm$ with the one obtained in the same region with
the slit alone (red and black curves, respectively), as well as in
the region immediately outside the slit exit (blue and green
curves). In addition, Fig. 2(d) shows the pattern of ${\bf E}({\bf
r})$ for the same configuration as Fig. 2(a). This illustrates the
enhancement of the electric field on the particle surface, as well
as the interesting vortices described by its wavevector around it
due to multiple reflections with the upper surface of the slab. This
is in contrast with the case in which the particle is dielectric and
the excited MDR is a whispering gallery mode (WGM), in which case
the field is mainly confined inside the particle, and exponentially
decays outside \cite{Chang}. In this latter case, the cylinder
reflectivity is much lower and the multiple reflections are much
weaker. This high reflectivity of the plasmonic cylinder produces
not only a high concentration of transmitted energy around its
surface, but also outside it. In particular, at the slit exit,
(compare Fig. 1(a) and Fig. 2(a)). This is related to a relatively
lower reflectivity at the slit entrance, and in fact, already this
lower reflectivity constitutes another signature of a larger
transmission into the space where the particle is. Hence, this
cylinder acts as an extractor of transmittance upon the slit (see
Figs. 2(a) and 2(d)).

\begin{figure}[htbp]
\centering
\includegraphics[width=16cm]{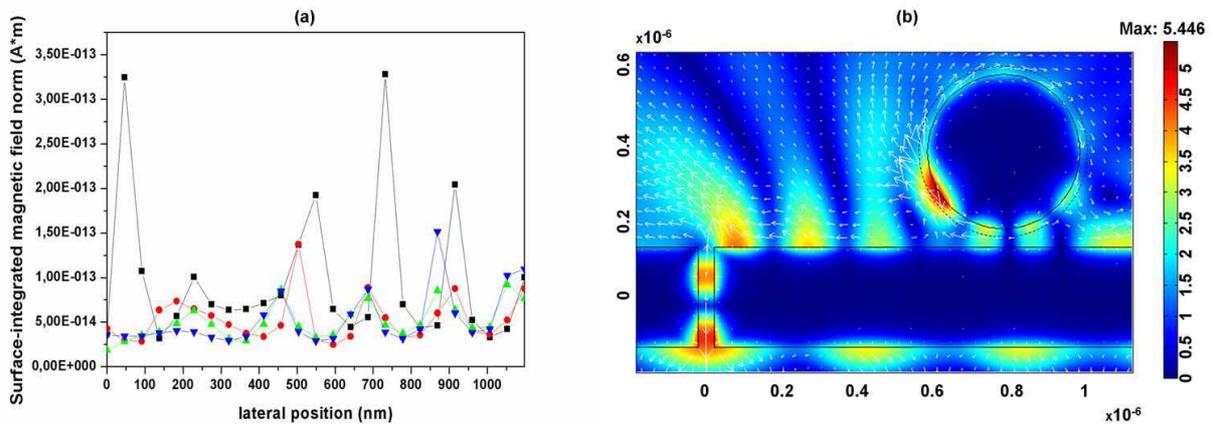}
\caption{(a) Variation of the magnetic field concentration $|{\bf
H}({\bf r})|$ (in $A/m$ units) on the cylinder surface vs. its
horizontal separation (in {\it nm}) from the slit exit. The particle
is illuminated at $\lambda= 366nm$ (i.e. near the nanocylinder
$LSP_{51}$ resonance). The black, red, green and blue curves stand
for vertical distances between the particle surface and the slab
equal to $\lambda/8$, $\lambda/4$, $3\lambda/8$ and $\lambda/2$,
respectively. (b) Map of $|{\bf H}({\bf r})|$ (color) and $<{\bf
S}({\bf r})>$) ($J/(m^{2}\cdot s)$) (arrows) when the cylinder is at
a vertical and horizontal distance from the slit exit: $\lambda/8$
and $17\lambda/8$, respectively, and illuminated at $\lambda=
364.7nm$ (p - polarization, near the $LSP_{51}$). Maximum arrow
length $\approx 26.09 KeV/(nm^{2}\cdot s)$, minimum arrow length
$\approx 0eV/(nm^{2}\cdot s)$. Calculations in Fig. 3(a) have been
made like in Figs. 2(b) and 2(c).}
\end{figure}

Figure 3(a) shows the behavior of both the magnetic field and the
energy flow with a lateral separation of the particle from the slit
exit. Now, this asymmetry of its location produces that the standing
LSP pattern on the cylinder surface disappears in its upper side,
showing the energy circulation around its surface. Also, a standing
wave is formed between the aperture and the reflecting cylinder
which projects energy into the upper space. Again, the
transmittivity of the slit is now much larger than when it is alone.

\subsection{Other sets of plasmonic nanocylinders}

\begin{figure}[htbp]
\centering
\includegraphics[width=16cm]{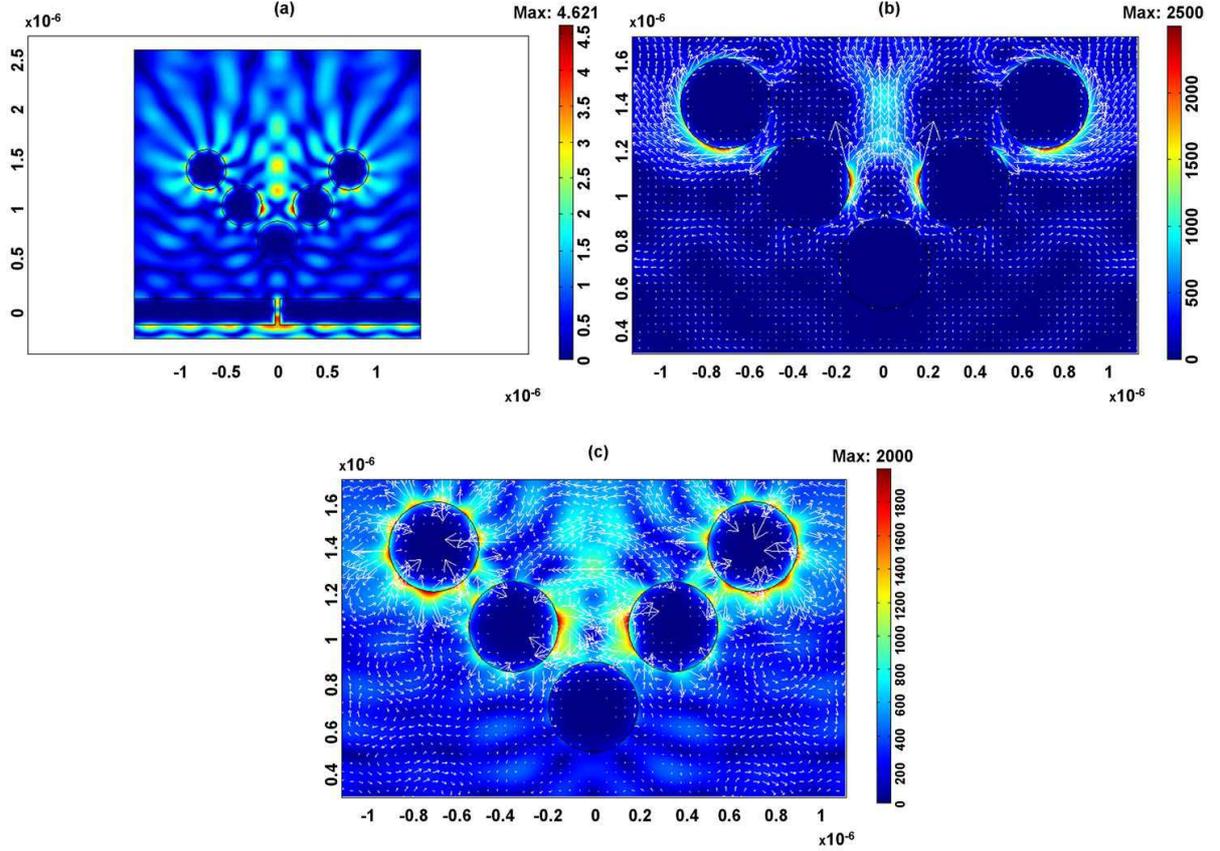}
\caption{(a) Magnetic field modulus $|{\bf H}({\bf r})|$ (in {\it
A/m}) spatial distribution for five $Ag$ cylinders, (radius $R=
200nm$, refractive index $n= 0.186+i1.610$), disposed in a
bifurcated chain, (bifurcation angle $\theta= 45^{\circ}$, distance
between cylinder surfaces $d_{cc}= 100nm$), in front of the slit of
Figs. 1(a) - (d) illuminated at $\lambda= 364.7nm$ (p -
polarization, near the $LSP_{51}$ resonance). The vertical distance
between the bottom cylinder surface and the exit plane of the slit
is $\lambda_{0}$, ($\lambda_{0}= 366nm$ near the $LSP_{51}$
resonance of an isolated cylinder); (b) Time averaged energy flow
$<{\bf S}({\bf r})>$ (arrows in $J/(m^{2}\cdot s$ units), maximum
arrow length $\approx 15.60 KeV/(nm^{2}\cdot s)$, minimum arrow
length $\approx 0eV/(nm^{2}\cdot s)$) in the same configuration as
in Fig. 4(a); (c) Electric field ${\bf E}({\bf r})$ distribution,
both norm (colors in {\it V/m} units) and directions (arrows) are
displayed in the same configuration as in Fig. 4(a).}
\end{figure}

In order to test light transmission and concentration by means of
aperture MDR - particle LSP interaction and through adjacent
cylinder LSP coupling, Figs. 4(a) - (c) show $|{\bf H}({\bf r)}|$,
$<{\bf S}({\bf r})>$ and ${\bf E}({\bf r})$ for a bifurcated chain
of Ag cylinders near a slit. The  distance between the slit and the
cylinders has been chosen so as to optimize the extraction of energy
by the particles through the aperture. Light transport through the
chain up to the upper particles is now obtained as a concentration
of energy spread in the region of the four upper cylinders as shown
in Figs. 4(a) - (c) with, of course, a lower intensity concentration
in the aperture exit zone. (compare $|<{\bf S}({\bf r})>|=
766.86eV/(nm^{2}\cdot s)$ in the area inside a rectangular monitor
appropriately scaled and equivalent to that shown in Fig. 2(a)) over
the slit exit when no particles are present against $|<{\bf S}({\bf
r})>|= 10698.50eV/(nm^{2}\cdot s)$ averaged over the same monitor in
the configuration of Figs. 4(a) - (c)). Notice that now, as before,
the optimum resonant wavelength $\lambda= 364.7nm$ for this combined
system of particles and slit is again red - shifted with respect to
the individual elements ($\lambda= 359.4nm$). The field transmission
through the particle chains is accomplished by both the propagating
waves surrounding the set and a field hopping process between
neighbor particles. An appropriate choice of set parameters and
illumination, allows one to select the particle of the set with the
most enhanced LSP field intensity on its surface.

\begin{figure}[htbp]
\centering
\includegraphics[width=16cm]{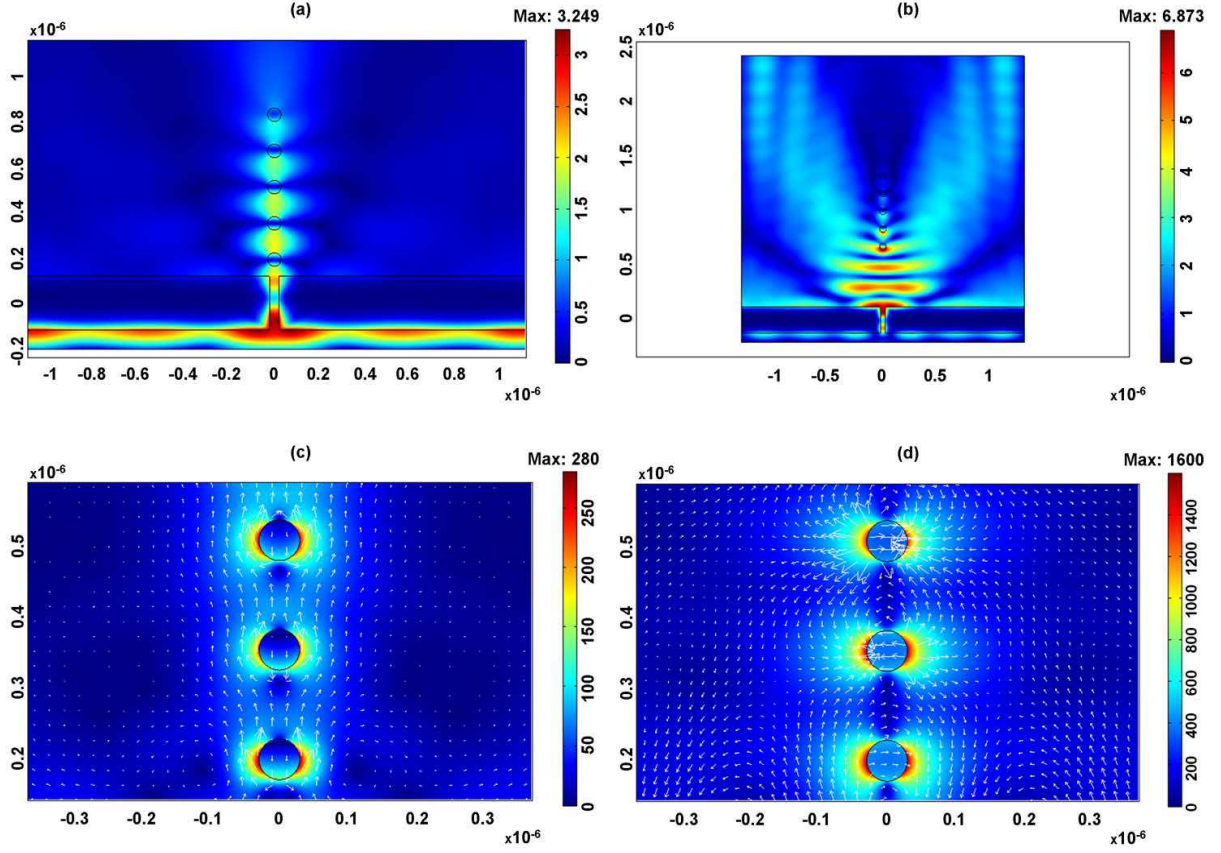}
\caption{(a) Magnetic field modulus $|{\bf H}({\bf r})|$ ({\it A/m})
distribution for five $Ag$ cylinders (radius $R= 30nm$, refractive
index $n= 0.173+i1.95$) of a linear chain, (distance between
cylinder surfaces $d_{cc}= 100nm$), in front of a slit in a W slab,
(slit width $d= 39.59nm$, slab width $D= 2610nm$, slab thickness $h=
237.55nm$, refractive index $n= 3.39+i2.41$) illuminated at
$\lambda= 400nm$ (p - polarization). The vertical distance between
the bottom cylinder surface and the exit plane of the slit is
$\lambda_{0}/8$ ( $\lambda_{0}= 346nm$ is a wavelength near the
$LSP_{11}$ of an isolated cylinder); (b) The same configuration as
in Fig. 4(a), changing the distance from the slit to
$3\lambda_{0}/2$ ($\lambda_{0}= 346nm$), illuminated at $\lambda=
349.3nm$ (p - polarization); (c) Detail of the time averaged energy
flow $<{\bf S}({\bf r})>$ (in $J/m^{2}\cdot s$) in the first three
cylinders of the chain for the arrangement of Fig. 4(a), both the
norm (colors) and directions are shown, (maximum arrow length
$\approx 1.75 KeV/(nm^{2}\cdot s)$, minimum arrow length $\approx
0eV/(nm^{2}\cdot s)$). (d) Detail of the electric field ${\bf
E}({\bf r})$ in the first three cylinders of the chain for the case
of Fig. 4(a), both the norm (colors in {\it V/m}) and directions
(arrows) are shown.}
\end{figure}

Chains of smaller size nanocylinders  at distance $d_{cc}$ from each
other, show dipolar plasmon coupling between neighbors, [see $|{\bf
H}({\bf r})|$ in Figs. 5(a) and 5(b)],  both near the exit plane of
the slit and at a larger distance from it. A detail of the
transmission in the first three cylinders of such a chain can be
seen in Figs. 5(c) - (d), where the energy flow and the electric
field are shown. The latter quantity exhibits on the surface of each
cylinder a typical dipolar LSP distribution, [see Fig. 5(d)]. This
transmitted  light through the particles does not take place by
tunneling, but by a hopping mechanism, like in dielectric molecular
photonic crystal rows \cite{Vigneron}, due to a $d_{cc}^{-2}$
dipolar interaction (where $d_{cc}$ is the distance between cylinder
surfaces) with frequency splitting of the single plasmonic cylinder
spectral line, and a redshift of its extinction peak. This coupling
diminishes the energy of the ensemble in the bonding state for this
configuration (parallel dipoles) \cite{Maier}. The field confinement
along the chain line and between cylinders in the case in which they
are close to the slit [cf. Figs. 5(a), (c) and (d)] contrasts with
the large diffraction occurring when their distance to the slab is
large [cf. Fig. 5(b)], this latter case also showing a strong
standing wave below the chain. The energy flow $|<{\bf S}>|$ in the
area immediately outside the slit exit is reduced with respect to
the case of the slit alone in the case of Fig. 5(a) (compare for
Fig. 5(a) $192.48eV/(nm^2\cdot s)$ against $1536.35eV/(nm^2\cdot s)$
for the same slit alone) however, it is enhanced in the case of Fig.
5(b) ($24674.90eV/(nm^2\cdot s)$ compared with $6531.39eV/(nm^2\cdot
s$) for the same slit alone. These numbers were obtained by
averaging in a suitable rectangular monitor at the slit exit.
Nevertheless, both linear configurations render a good response
regarding to the averaged energy flow concentrated around the
cylinders. (See the corresponding low reflected energy in Fig.
5(b)). The linear chain near the slit of Fig. 5(a) achieves an
averaged $|<{\bf S}>|$ of $1100eV/(nm^2\cdot s)$ against
$188.87eV/(nm^2\cdot s)$ in the same area when the slit transmits
without the cylinders.

\begin{figure}[htbp]
\centering
\includegraphics[width=8cm]{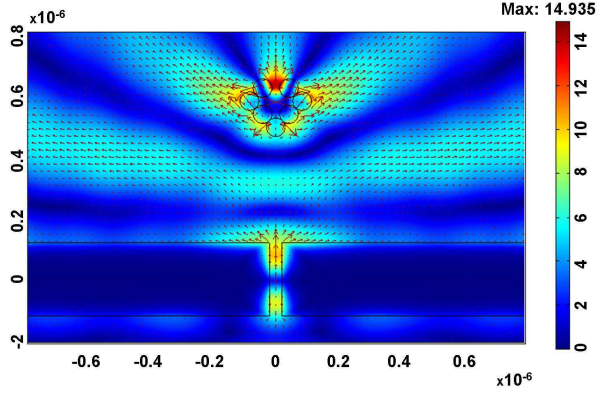}
\caption{Magnetic field modulus $|{\bf H}({\bf r})|$ (colors in {\it
A/m} units) and time averaged energy flow $<{\bf S}({\bf r})>$
(arrows in $J/m^{2}\cdot s$ units, maximum arrow length $\approx
1.52\cdot 10^{5} KeV/(nm^{2}\cdot s)$, minimum arrow length $\approx
0eV/(nm^{2}\cdot s)$) concentrated on the surfaces of seven $Ag$
cylinders (radius $R= 30nm$, refractive index $n= 0.186+i1.61$) in
bifurcation, (bifurcation angle $\theta_{1}= 45^{\circ}$, distance
between cylinder surfaces $d_{cc}= 0nm$), with elbow, (elbow angle
$\theta_{2}= 45^{\circ}$), in front of a slit in a W slab (slit
width $d= 39.59nm$, slab width $D= 2610nm$, slab thickness $h=
237.55nm$, refractive index $n= 3.39+i2.66$), illuminated at
$\lambda= 364.7nm$ (p-polarization) which is near the $LSP_{11}$ of
each cylinder. The vertical distance between the bottom cylinder
surface and the exit plane of the slit is $d_{lc}= \lambda_{0}=
346nm$, near the $LSP_{11}$ of the isolated cylinder.}
\end{figure}

Other distributions considered include those of bifurcations with
elbows of nanoparticles at subwavelength distance $d_{cc}$ from each
other, as illustrated in Figure 6, which shows ($|{\bf H}({\bf r})|$
in color and $<{\bf S}({\bf r})>$ in arrows. Now the dipolar
interaction is of order $d_{cc}^{-3}$ since it takes place in the
near field. The enhancement of transmitted light on top of the set
is now quite sharp as the cylinders approach each other (not shown).
This set presents the bottom cylinder placed at a distance from the
slit that optimizes the energy concentration both  on the top
particles and at the aperture exit. Most light passes through the
aperture and reaches the cylinder set, although the radiation
pattern reveals coupling between cylinders and slab. Notice the low
value of $|{\bf H(r)}|$ in the reflection region below the slab in
Fig. 6. The standing wave between the slab and the cylinders now
embraces the set in one of the maxima, and makes it to strongly emit
upwards along three main directions, hence exhibiting a
nanoantenna-like behavior. (Radiation directions can be controlled
by changing the configuration at the chosen wavelength). Again, this
effect is more pronounced with the cylinders at a certain distance
from the slit, as shown in Fig. 6, than when they are very close to
it. In either case, however, this intensity, strongly transmits even
into the aperture exit ($42216.07eV/(nm^{2}\cdot s)$ for Fig. 6
against $2783.01eV/(nm^{2}\cdot s)$ for the same slit alone), (these
values obtained by an average over the area of a suitable
rectangular monitor at the slit exit, although most energy is now
transmitted to the cylinders, and associated with a very small
reflection below the slab. This process of transmission up to the
top of the set is much more efficient than that of transmission by
the excitation of the morphological resonance of the slit alone,
which was shown in Fig. 2(c).

\subsection{A metallic photonic crystal in front of an array of slits}

Aperture supertransmission is further enhanced by periodically
repeating it in the slab \cite{Ebbesen98} - \cite{Garc�a07}.
Accordingly, we next arrange chains of metallic particles, each
placed in front of a slit of such an array. Introducing some
distance between cylinders in each chain, we build in this form a
metallic photonic crystal (PC).



\begin{figure}[htbp]
\centering
\includegraphics[width=16cm]{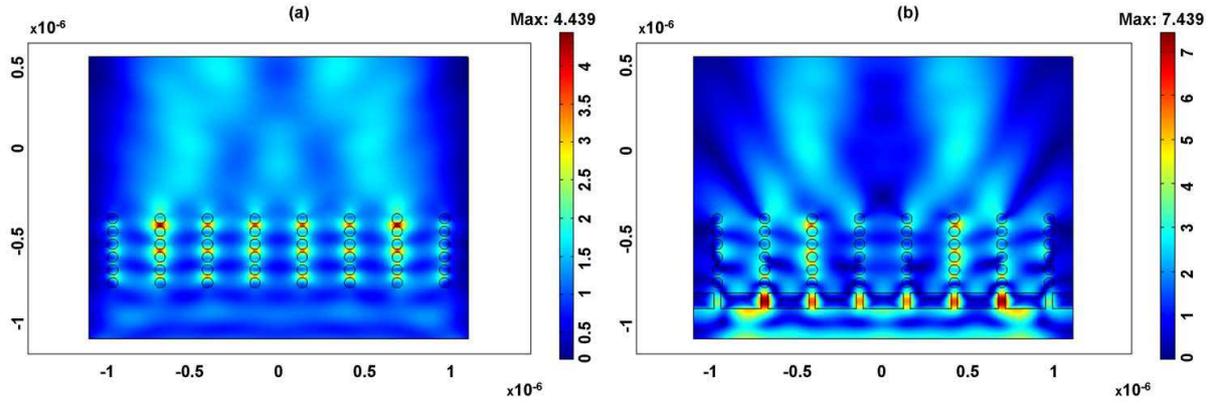}
\caption{(a) Magnetic field modulus $|{\bf H}({\bf r})|$ ({\it A/m})
distribution in a photonic crystal (horizontal period $a_{x}=
275.50nm$, vertical period $a_{y}= 75nm$) formed with 68 $Ag$
cylinders (radius $R= 30nm$, refractive index $n= 0.173+i1.95$) and
illuminated at $\lambda= 400nm$ (p - polarization, the $LSP_{11}$
excited is near that of the isolated particle which would appear at
$\lambda\approx 346nm$); (b) Magnetic field modulus $|{\bf H}({\bf
r})|$ distribution in the $Ag$ photonic crystal of Fig. 7(a), this
time in front of a grating of slits practiced in a $W$ slab (period
$P= 275.50nm$, slit width $d= 39.36nm$, slab width $D= 8P$, slab
thickness $h= 94.46nm$, refractive index $n= 3.39+i2.41$),
illuminated at $\lambda= 400nm$ (i.e. near the $LSP_{11}$ of an
isolated particle), (p - polarization). The distance between the
cylinder surfaces of the first row and the exit plane of the slits
is 22.5nm (fitted to get the best response).}
\end{figure}

Figures 7(a) and 7(b), corresponding to the PC alone and to the PC
in front of the slit array, respectively, show the first step of our
approach. At the chosen illumination wavelength $\lambda= 400nm$ the
slit grating is supertransmitting and so does the array because this
wavelength is both close to a Rayleigh resonance of the slit array
and near the $LSP_{11}$ of the PC cylinders. As a result, a strong
concentration of field in the vertical rows of the PC appears. This
is again due to a dipolar interaction between neighbor particles of
the same vertical row and between adjacent rows as discussed in
connection with Figs. 5(a) - (d). The enhancement of the field is
alternating between particle gaps in each row, at difference with
the case of a single chain, shown in Figs. 5(a) - (d). The
qualitative aspect of the distribution of Figs. 7(a) and 7(b) is
similar, except for the collimation effect produced by the slit
grating, and the slight weakening of the energy enhancement between
particles due to intensity concentration within the slits, shown in
Fig. 7(b). These results in Fig. 7(b) contrast with those in which
the incident wavelength is out of resonance of the cylinders, then
light passes through the PC with scarce interaction with the
cylinders (not shown here).

\begin{figure}[htbp]
\centering
\includegraphics[width=16cm]{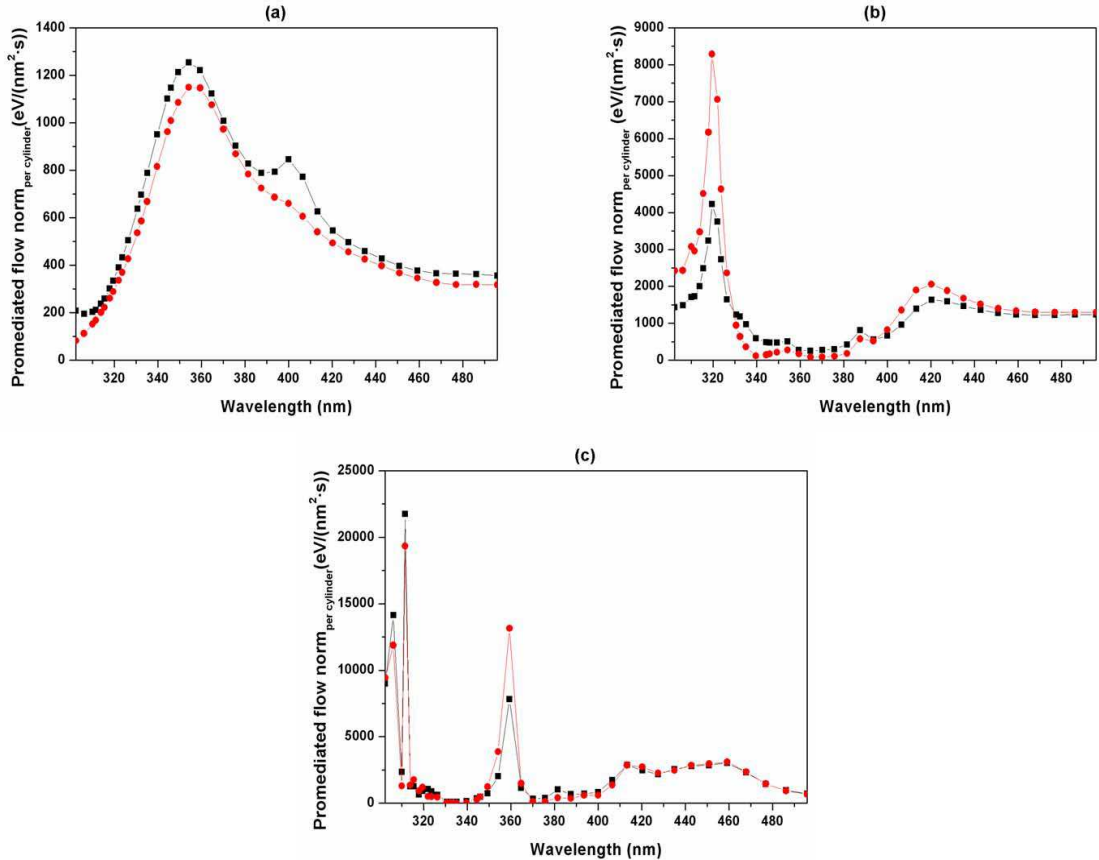}
\caption{(a) Norm $|<{\bf S}>|$ in $eV/(nm^{2}\cdot s)$ vs.
wavelength $\lambda$ (in {\it nm}) transmitted by the $W$ array of
slits alone of Fig. 7(b). (b) Average energy flow norm $|<{\bf S}>|$
in $eV/(nm^{2}\cdot s)$ units vs. wavelength passing through the PC
of $Ag$ cylinders of radius $R= 30nm$, [horizontal period $a_{x}=
275.50nm$, vertical period $a_{y}= 160nm$, see also Figs. 9(a) -
(c)]. (c) Norm ($|<{\bf S}>|$) vs. wavelength for the same PC, now
in front of the $W$ grating of Figure 8(a). The black squared and
red circled curves stand for the norm of energy flow averaged over
each square ($120 \times 120 nm^{2}$) circumscribed to each
cylind.er section and over each rectangular strip ($120 \times
920nm^{2}$) circumscribed to each PC vertical row, respectively. In
the case of Figure 8(a), these circles are imaginary and coincide
with the cylinders sections of Figures 8(b) and 8(c), (see also
Figs. 9(a) - (c)).
}
\end{figure}

In the second step of our analysis, the distance between cylinders
in each vertical chain is incremented to a value comparable to that
of the horizontal distance between chains. Figures 8(a) - (c)
correspond to the responses of the grating alone, this new PC alone
and the combination of both, respectively. Figure 8(b) shows an
effective bandgap in the $XM$ direction of transmission in PC
reciprocal space (upwards direction in Figs. 7(a) - (b) and Figs.
9(a) - (c)). Nevertheless, an enhancement of transmission along this
direction is achieved in the whole range studied (compare the values
of Fig. 8(b) to those of Fig. 8(c)). Furthermore, a transmission
peak rises for the PC/grating arrangement, in the range where the
gap was for the PC alone (see Figs. 8(c) and 8(b)), the two last
effects being due to the presence of the grating. Note the match in
wavelength between the transmission peak of the grating alone in
Fig. 8(a) and that mentioned above for the combination of PC and
grating in Fig. 8(c).

\begin{figure}[htbp]
\centering
\includegraphics[width=16cm]{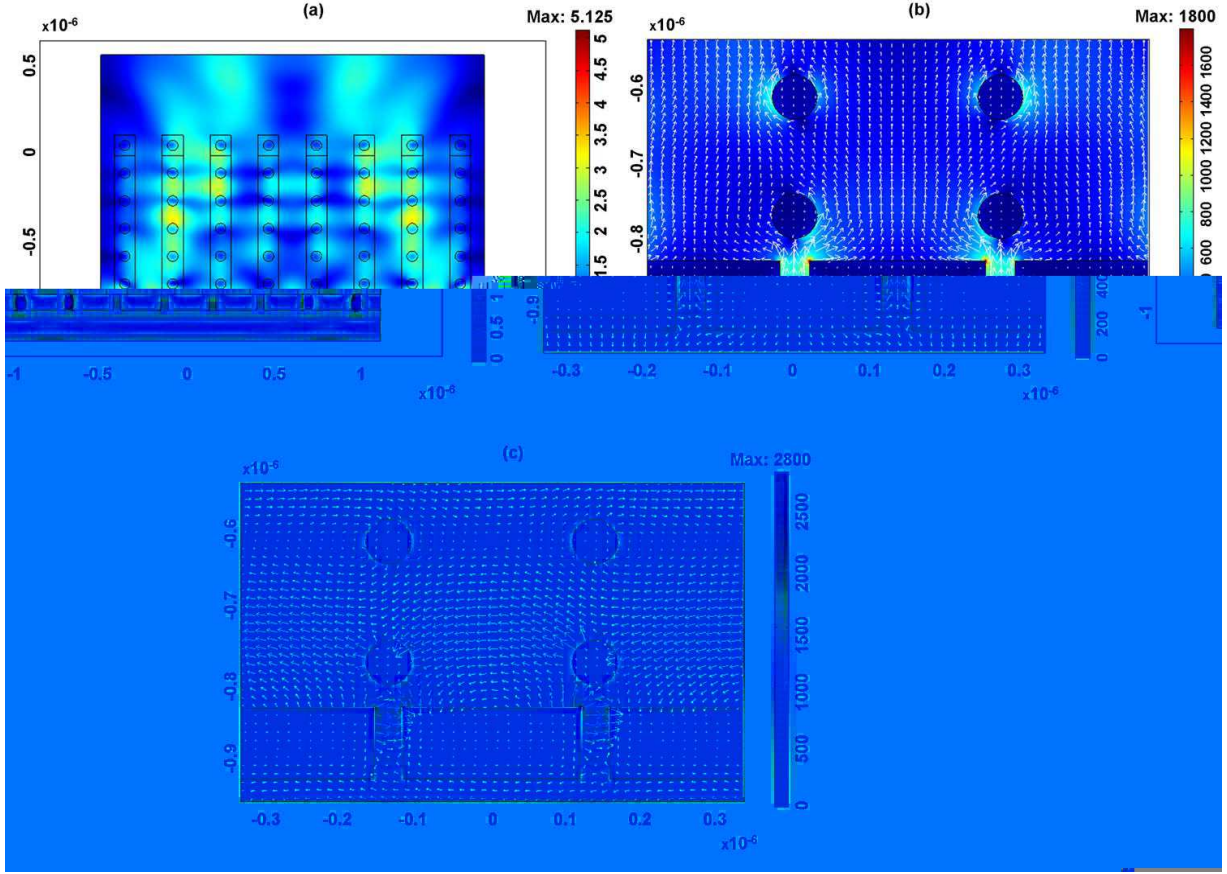}
\caption{(a) Map of the magnetic field norm $|{\bf H}({\bf r})|$ (in
{\it A/m} units) for the $Ag$ cylinder photonic crystal, studied in
Fig. 8(b), placed in front of the exit of the $W$ array of Figs.
7(b) and 8(a), illuminated at $\lambda= 450nm$ (see Fig. 8(c)). The
distance between the first horizontal row lower edges and the slab
is 22.5nm. (b) Detail of the time averaged energy flow ${\bf S}({\bf
r})$ (in $J/m^{2}\cdot s$ units), showing both its norm (color) and
directions (arrows), maximum arrow length $\approx 11.23
KeV/(nm^{2}\cdot s)$, minimum arrow length $\approx 0eV/(nm^{2}\cdot
s)$). (c) Detail of the electric field ${\bf E}({\bf r})$ (in {\it
V/m} units), both norm (color) and directions (arrows) are shown.}
\end{figure}

By contrast to Fig. 7(b), when the slit grating is added to this PC,
and illumination is out of both the LSP resonance and the PC
effective gap, (see Fig. 8(b)), one again observes an enhancement of
transmission into the PC, a strong intensity concentration in the
slits, and a relatively low reflection below the slab (as example,
that shown in Figs. 9(a) - (c)), even though now this reflection is
larger, relative to the transmittivity of the slits, than that shown
for resonantly excited nanoparticle sets in front of one slit
(compare with Figs. 5(a) - (d) and 6). On the other hand, the
presence of the slits modifies the gap of the PC which, as shown in
Fig. 8(c) has a transmission peak near $\lambda= 360nm$. This is a
new feature of the effect of the slits upon the transmittivity of
the PC.

\newpage

\section{Conclusion}

We have presented a theoretical and simulation study which shows how
the excitation of localized surface plasmons of nanoparticles in
front of subwavelength apertures, ($Ag$ cylinders and $W$ slabs have
been used) under polarization that excites the aperture propagating
modes (p - polarization in 2D), may enhance the transmission. Sets
of plasmonic cylinders exhibit coupling and propagation through
their elements. There is a specific distance for each of such
cylinder - slit configurations that optimizes the transmitted energy
passing into the particles. This suggests a control of light
transmission via plasmonic particles. Also, we observe that it is
possible to fitting particle set parameters and illumination in such
a way that the transmitted intensity is concentrated in certain
cylinders when the stationary regime of propagation has been
reached.

The case of a metallic photonic crystal ($Ag$ PC) in front of a
metallic slit array shows the effects  of effective bandgaps on the
interparticle LSP transmission, and  the enhancement of the
transmittivity of the grating due to the excitation of LSPs in the
metallic PC. All these results are also expected with 3D particles
in front of apertures with any geometry, in particular subwavelength
holes, and have a potential for controlling transmitted near fields
at the nanoscale.

\section*{Acknowledgements}
Work supported by the Spanish MEC through FIS2009-13430-C02-C01 and
Consolider NanoLight (CSD2007-00046) research grants, FJVV is
supported by the last grant.





\end{document}